\documentclass[fleqn,10pt]{wlscirep}
\usepackage[utf8]{inputenc}
\usepackage[T1]{fontenc}
\usepackage{multirow}

\usepackage{makecell}

 \title{A Robust Ensemble Algorithm for Ischemic Stroke Lesion Segmentation: Generalizability and Clinical Utility Beyond the ISLES Challenge}

\author[1,2,3*]{Ezequiel de la Rosa}
\author[4,5,6]{Mauricio Reyes}
\author[7,8]{Sook-Lei Liew}
\author[7]{Alexandre Hutton}
\author[9,10]{Roland Wiest}
\author[9,10]{Johannes Kaesmacher}
\author[11]{Uta Hanning}
\author[9]{Arsany Hakim}
\author[9]{Richard Zubal}
\author[9,10]{Waldo Valenzuela}
\author[3]{David Robben}
\author[3]{Diana M. Sima}
\author[3]{Vincenzo Anania}
\author[3]{Arne Brys}
\author[12]{James A. Meakin}
\author[12]{Anne Mickan}
\author[11]{Gabriel Broocks}
\author[11]{Christian Heitkamp}
\author[13]{Shengbo Gao}
\author[14]{Kongming Liang}
\author[14]{Ziji Zhang}
\author[15]{Md Mahfuzur Rahman Siddiquee}
\author[16]{Andriy Myronenko}
\author[17]{Pooya Ashtari}
\author[17]{Sabine Van Huffel}
\author[18]{Hyun-su Jeong}
\author[18]{Chi-ho Yoon}
\author[18]{Chulhong Kim}
\author[19]{Jiayu Huo}
\author[19]{Sebastien Ourselin}
\author[19]{Rachel Sparks}
\author[20]{Albert Clèrigues}
\author[20]{Arnau Oliver}
\author[20]{Xavier Lladó}
\author[21]{Liam Chalcroft}
\author[22]{Ioannis Pappas}
\author[23]{Jeroen Bertels}
\author[23]{Ewout Heylen}
\author[24]{Juliette Moreau}
\author[24]{Nima Hatami}
\author[24]{Carole Frindel}
\author[25]{Abdul Qayyum}
\author[26]{Moona Mazher}
\author[27]{Domenec Puig}
\author[28]{Shao-Chieh Lin}
\author[28]{Chun-Jung Juan}
\author[29]{Tianxi Hu}
\author[29]{Lyndon Boone}
\author[29,30]{Maged Goubran}
\author[31]{Yi-Jui Liu}
\author[32,33]{Susanne Wegener}
\author[34,2,35,36]{Florian Kofler}
\author[2,36]{Ivan Ezhov}
\author[2,36]{Suprosanna Shit}
\author[35]{Moritz R. Hernandez Petzsche}
\author[1]{Bjoern Menze}
\author[35,36,+]{Jan S. Kirschke}
\author[35,36,+]{Benedikt Wiestler}

\affil[1]{Department of Quantitative Biomedicine, University of Zurich, Zurich, Switzerland.}
\affil[2]{Department of Informatics, Technical University Munich, Germany.}
\affil[3]{ico\textbf{metrix}, Leuven, Belgium.}
\affil[4]{ARTORG Center for Biomedical Research, University of Bern, Bern, Switzerland.}
\affil[5]{Department of Radiation Oncology, University Hospital Bern, University of Bern.}
\affil[6]{University of Bern, Bern, Switzerland.}
\affil[7]{Chan Division of Occupational Science and Occupational Therapy, University of Southern California, Los Angeles, CA United States.}
\affil[8]{Stevens Neuroimaging and Informatics Institute, Department of Neurology, Keck School of Medicine, University of Southern California.}
\affil[9]{Support Center of Advanced Neuroimaging (SCAN), University Institute of Diagnostic and Interventional Neuroradiology, Inselspital, Bern, Switzerland.}
\affil[10]{University Institute of Diagnostic and Interventional Neuroradiology, University Hospital Bern, Inselspital, University of Bern, Bern, Switzerland.}
\affil[11]{Department of Diagnostic and Interventional neuroradiology, University Medical Center Hamburg-Eppendorf, Hamburg, Germany.}
\affil[12]{Department of Medical Imaging, Radboud University Medical Center, Institute for Health Sciences, Nijmegen, The Netherlands.}
\affil[13]{Deepwise AI Lab, Beijing, China.}
\affil[14]{Beijing University of Posts and Telecommunications, Bejing, China.}
\affil[15]{School of Computing and Augmented Intelligence, Arizona State University, Tempe, AZ, USA.}
\affil[16]{NVIDIA, Santa Clara, CA, USA.}
\affil[17]{Department of Electrical Engineering (ESAT), STADIUS Center for Dynamical Systems, Signal Processing, and Data Analytics, KU Leuven, Leuven, Belgium.}
\affil[18]{Graduate School of Artificial Intelligence (GSAI), Department of Electrical Engineering, Convergence IT Engineering, Mechanical Engineering, Medical Science and Engineering, and Medical Device Innovation Center, Pohang University of Science and Technology (POSTECH), Pohang, South Korea.}
\affil[19]{School of Biomedical Engineering \& Imaging Sciences, King’s College London, UK.}
\affil[20]{Institute of Computer Vision and Robotics, University of Girona, Spain.}
\affil[21]{Wellcome Centre for Human Neuroimaging, University College London, London, UK.}
\affil[22]{Laboratory of Neuro Imaging, Stevens Institute for Neuroimaging and Informatics, Keck School of Medicine, University of Southern California, Los Angeles, United States.}
\affil[23]{Department of Electrical Engineering (ESAT), Processing Speech and Images (PSI), KU Leuven, Leuven, Belgium.}
\affil[24]{CREATIS, Université Lyon1, CNRS UMR5220, INSERM U1206, INSA-Lyon, 69621 Villeurbanne, France.}
\affil[25]{National Heart and Lung Institute, Faculty of Medicine, Imperial College London, UK.}
\affil[26]{Centre for Medical Image Computing, Department of Computer Science, University College London, UK.}
\affil[27]{Department of Computer Engineering and Mathematics, University Rovira I Virgili, Spain.}
\affil[28]{Department of Medical Imaging, China Medical University Hsinchu Hospital, Hsinchu, Taiwan, Republic of China.}
\affil[29]{Department of Medical Biophysics, University of Toronto, Toronto, Canada.}
\affil[30]{Hurvitz Brain Sciences Research Program, Sunnybrook Research Institute, Toronto, Canada.}
\affil[31]{Department of Automatic Control Engineering, Feng Chia University, Taichung, Taiwan, Republic of China.} 
\affil[32]{Department of Neurology, University Hospital of Zurich, Zurich, Switzerland.} 
\affil[33]{University of Zurich, Zurich, Switzerland.} 
\affil[34]{Helmholtz AI, Helmholtz Munich, Neuherberg, Germany.} 
\affil[35]{Department of Diagnostic and Interventional Neuroradiology, School of Medicine, Klinikum rechts der Isar, Technical University of Munich, Germany.}
\affil[36]{TranslaTUM, Center for Translational Cancer Research, Technical University of Munich, Germany.}

\affil[*]{Corresponding author: ezequieldlrosa@gmail.com}
\affil[+]{These authors contributed equally.}

\begin{abstract}
Diffusion-weighted MRI (DWI) is essential for stroke diagnosis, treatment decisions, and prognosis. However, image and disease variability hinder the development of generalizable AI algorithms with clinical value. We address this gap by presenting a novel ensemble algorithm derived from the 2022 Ischemic Stroke Lesion Segmentation (ISLES) challenge.

ISLES'22 provided 400 patient scans with ischemic stroke from various medical centers, facilitating the development of a wide range of cutting-edge segmentation algorithms by the research community. By assessing them against a hidden test set, we identified strengths, weaknesses, and potential biases. Through collaboration with leading teams, we combined top-performing algorithms into an ensemble model that overcomes the limitations of individual solutions.

Our ensemble model combines the individual algorithms' strengths and achieved superior ischemic lesion detection and segmentation accuracy (median Dice score: 0.82, median lesion-wise F1 score: 0.86) on our internal test set compared to individual algorithms. This accuracy generalized well across diverse image and disease variables. Furthermore, the model excelled in extracting clinical biomarkers like lesion types and affected vascular territories. Notably, in a Turing-like test, neuroradiologists consistently preferred the algorithm's segmentations over manual expert efforts, highlighting increased comprehensiveness and precision.

Validation using a real-world external dataset (N=1686) confirmed the model's generalizability (median Dice score: 0.82, median lesion-wise F1 score: 0.86). The algorithm's outputs also demonstrated strong correlations with clinical scores (admission NIHSS and 90-day mRS) on par with or exceeding expert-derived results, underlining its clinical relevance.

This study offers two key findings. First, we present an ensemble algorithm that detects and segments ischemic stroke lesions on DWI across diverse scenarios on par with expert (neuro)radiologists. Second, we show the potential for biomedical challenge outputs to extend beyond the challenge's initial objectives, demonstrating their real-world clinical applicability. Our publicly available algorithm (\url{https://github.com/Tabrisrei/ISLES22_Ensemble}) has the potential to significantly improve stroke diagnosis and patient care.
\end{abstract}

\begin{document}

\flushbottom
\maketitle
\thispagestyle{empty}

\noindent 

\section*{Introduction}

Brain imaging is crucial to evaluate tissue viability and fate in ischemic stroke. Magnetic resonance imaging (MRI) supports physicians across diverse stages of the disease. It helps define the optimal reperfusion treatment, unveils the stroke etiology, and sheds light on prognostic clinical outcomes \cite{czap2021overview}. Diffusion-weighted imaging (DWI) is considered the current gold standard for imaging the ischemic core \cite{goyal2020challenging, giancardo2023segmentation}. Although imperfect, DWI is the only imaging technique reliably demonstrating parenchymal injury within minutes to hours from the stroke onset \cite{merino2010imaging}. Currently, deep learning algorithms are revolutionizing medical imaging, demonstrating unprecedented performance across multiple radiological tasks. Segmentation of ischemic stroke tissue using deep learning has been proposed in different works \cite{chen2017fully, federau2020improved, liu2021deep, alis2021inter, nazari2023predicting}. 
The complexity of the task lies in multiple sources of variability that involve image- (e.g., driven by center- and scanner-specific MRI acquisition differences, artifacts mimicking ischemic lesions \cite{roberts2003diffusion}, time-dependent DWI signaling \cite{merino2010imaging, chalela2007magnetic}, etc.), patient- (e.g., age \cite{fung2011mr}) and disease-specific characteristics (such as the subtype of stroke and its etiology \cite{kang2003association}). Little is known, however, about the real-world transferability potential of deep learning algorithms for ischemic stroke segmentation, their generalization towards diverse cohorts and image characteristics, and their ultimate clinical utility. 

Biomedical challenges are international competitions aiming to benchmark task-specific algorithms under controlled settings \cite{maier2018rankings}. The organization of medical image challenges has rapidly grown, enabling to tackle problems related to diverse organs, tasks (e.g., lesion detection or anatomy segmentation), and image modalities (such as MRI, CT, among others) \cite{menze2014multimodal, litjens2014evaluation, sekuboyina2021verse, sirinukunwattana2017gland, bilic2023liver}. Challenges are now considered a \emph{de facto} gold standard for algorithm comparison by the research community \cite{antonelli2022medical} and have also been adopted by the Radiological Society of North America (RSNA)\footnote{https://www.rsna.org/rsnai/ai-image-challenge}. Segmentation of stroke lesions from MRI has not been an exception, and the number of methods devised targeting this task considerably increased following the 2015 Ischemic Stroke Lesion Segmentation (ISLES) challenge \cite{maier2017isles}. ISLES'15 is considered a reference evaluation tool for the segmentation of brain ischemia. In the past few years, studies highlighting the strengths and weaknesses of challenge organization emerged, providing good implementation practices \cite{reinke2018exploit, maier2018rankings, maier2020bias}. Such initiatives considerably improved the quality of current challenges in terms of execution, interpretation, fairness, transparency, and reproducibility. 

Biomedical challenges, when properly designed, are powerful. They operate as international problem-solving sprints that involve leading researchers worldwide. Therefore, we take advantage of such an event to rapidly prototype and identify candidate ischemic stroke segmentation algorithms. We hypothesized that (1) a challenge might yield an algorithm or a strategy that reliably detects and segments brain ischemia under real-world, heterogeneous data scenarios, and (2) such an algorithm may generalize above the challenge context to real-world data, thus becoming relevant to downstream clinical analysis. We organized the ISLES'22 challenge during the 2022 International Conference on Medical Image Computing and Computer-Assisted Intervention (MICCAI) \cite{GC_isles, islesdocument} to test these hypotheses. ISLES'22 builds on top of the experience gained from the earlier ISLES'15, overcoming some of its drawbacks by adhering to current challenge standards \cite{reinke2018exploit, maier2020bias}, by using a standardized platform \cite{GC_isles} for the fair assessment of software solutions, and by including more than six times the number of patients than ISLES'15. This study describes the organization of the challenge, the development of a robust deep learning ensemble solution capable of generalizing to multiple axes of data variability, its validation using the largest available (external) stroke dataset, and, ultimately, an illustration of its practical utility in facilitating downstream clinical analysis on real-world data. The contribution of this study is two-fold. First, we introduce an algorithm capable of identifying and segmenting ischemic stroke in MRI at a level comparable to expert radiologists. Furthermore, through a Turing-like experiment, we found that the algorithm produces superior qualitative segmentations, which neuroradiologists prefer over manually delineated expert counterparts. Secondly, our investigation is pioneering in reporting a challenge-derived algorithm whose (clinical) utility and applicability transcend the primary challenge objectives, showcasing its usefulness for subsequent clinical analyses.

\section*{Results}
\subsection*{Challenge overview} 
\begin{figure}[hbt!]
\begin{center}
\includegraphics[width=16cm]{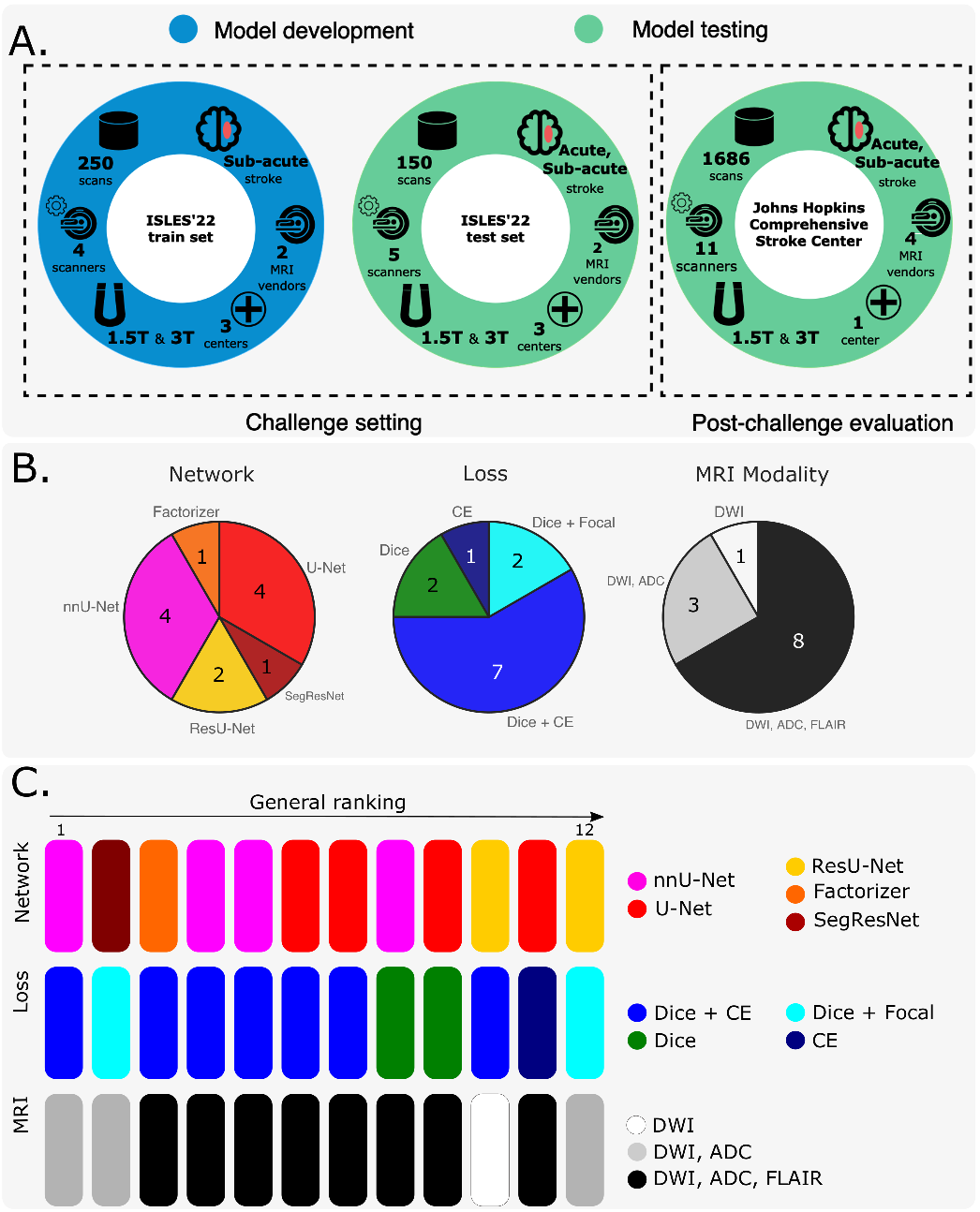}
\caption{Overview of the ISLES'22 challenge and post-challenge experimental design, including the developed algorithmic solutions. A) Challenge and post-challenge phases and datasets. B) Summary of algorithmic solutions stratified by network architecture, loss function, and input modalities. C) Challenge leaderboard stratified by network architecture, loss function, and input modalities. CE: Cross-entropy.}

\label{fig:cnn_summary}
\end{center}
\end{figure}

Figure \ref{fig:cnn_summary}(A) summarizes the ISLES'22 challenge. In the model development phase, teams utilized labeled datasets to create algorithmic solutions, which were evaluated using undisclosed data in the model testing phase. Differently from other imaging challenges, the challenge datasets were kept raw and un-processed to emulate a real-world setting, thus challenging participants to devise end-to-end algorithmic pipelines.  
Figure \ref{fig:cnn_summary}(A) also showcases an overview of the external dataset used to evaluate algorithmic performance in a post-challenge, `real-world' data setting. The ISLES'22 challenge was held during May-August 2022. By the challenge closing date, there were 476 registered participants and 325 dataset downloads. A total of 20 teams tested their algorithms on the remote server using the \emph{sanity-check} phase. Finally, we received 15 submissions to the challenge's final \emph{test-phase}, from which 12 fulfilled the ISLES'22-MICCAI requirements \cite{islesdocument}. The 12 participating teams submitted deep learning-based solutions. In Figure \ref{fig:cnn_summary}(B), a fingerprint of the methods is shown. nnUnet \cite{isensee2021nnu} and UNet-like \cite{ronneberger2015u, cciccek20163d} neural networks were the most prevalent architectures. The most frequently used loss function was a combination of Dice with categorical cross-entropy. Figure \ref{fig:cnn_summary}(C) shows a pictorial ranking of the algorithmic submissions. Most teams preferred to use all the available MRI modalities (DWI, apparent diffusion coefficient `ADC', and fluid-attenuated inversion recovery `FLAIR') as inputs to the model. It is worth observing that the top-3 ranked teams used diverse deep learning pipelines in terms of architecture, loss function, and model inputs.

\subsection*{Performance and ranking}
\begin{figure}[!b]
\begin{centering}
\includegraphics[width= 13cm]{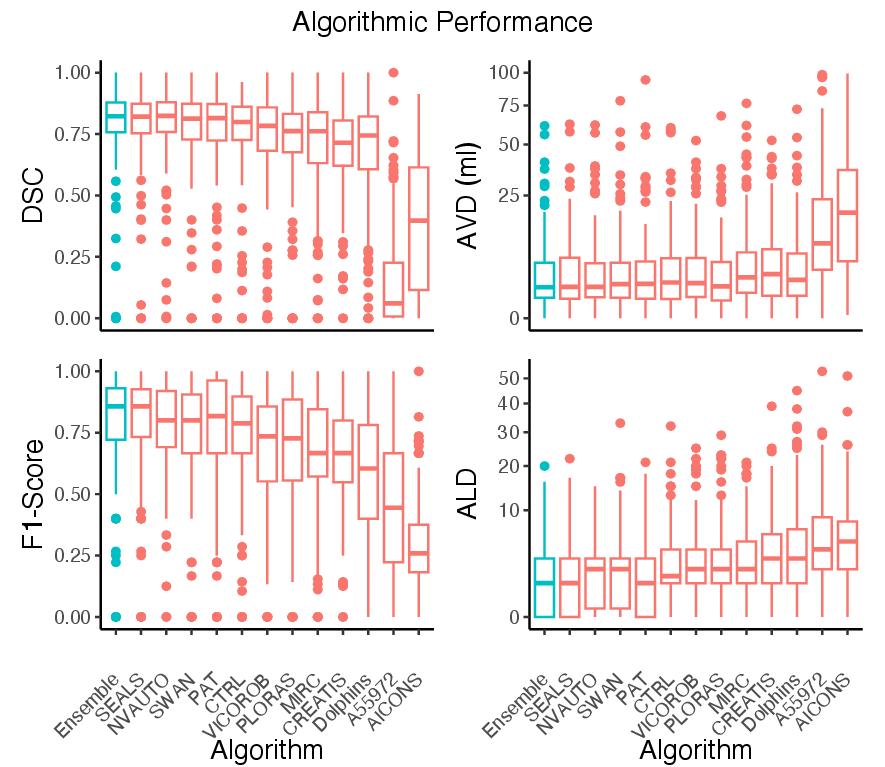}
\caption{Performance for the participating teams in the unseen test phase of the challenge. Teams are displayed in red and in decreasing order based on their final rank. DSC: Dice Similarity Coefficient; F1 score: lesion-wise F1 score; AVD: absolute volume difference; ALD: absolute lesion count difference. y-axis for AVD and ALD boxplots are displayed using a non-linear scale to enhance data visibility.}
\label{fig:T1-boxplots}
\end{centering}
\end{figure}

The final ISLES'22 leaderboard is shown in Table \ref{table:ranking}. In Figure \ref{fig:T1-boxplots}, boxplots with performance metrics are displayed. Further statistics about the challenge rankings obtained through a thousand bootstraps are available in Supplementary Materials S1-S2. While the winner of the challenge (team `SEALS') led the leaderboard in the detection metrics (lesion-wise F1 score and absolute lesion count difference), the second-ranked team (`NVAUTO') led the leaderboard in segmentation-derived metrics (Dice and absolute volume difference, see S1-S2). In the `rank then aggregate' ranking calculation (Table \ref{table:ranking} and Suppl. mat. S1), that compares methods case-wise, the teams SWAN and PAT achieved a shared third position, whereas in the complementary ranking calculation using the `aggregate then rank' approach (Suppl. mat. S2), they secured a third and fourth position, respectively.
In a post-challenge scenario, we created an ensemble algorithm that incorporates solutions from three top-ranked teams, namely SEALS, NVAUTO, and SWAN. We then recomputed the rankings, introducing the ensemble as a new participant. Notably, the ensemble algorithm achieves the top-ranking position with exceptional performance across all the evaluated metrics. Additional information regarding the ensemble's performance can be found in the summary of the thousand bootstraps experiment presented in Supplementary Material S3. Such analyses include results about ranking stability, ranking robustness to ranking methods, and inter-algorithm statistical tests.

\begin{table}[!t]
\begin{center}

\begin{tabular}{rlllll}
                  \\

\multicolumn{1}{c}{\textbf{Rank}} & \multicolumn{1}{c}{\textbf{Team}} & \multicolumn{1}{c}{\textbf{DSC $\uparrow$}} & \multicolumn{1}{c}{\textbf{AVD (ml) $\downarrow$}} & \multicolumn{1}{c}{\textbf{F1 $\uparrow$}} & \multicolumn{1}{c}{\textbf{ALD $\downarrow$}} \\
\hline 
*1                                & Ensemble                            & \textbf{0.82} $\pm$ 0.12                       & \textbf{1.59} $\pm$ 4.40                    & \textbf{0.86} $\pm$ 0.21                     & \textbf{1.00} $\pm$ 3.00\\ \hline 
1                                 & SEALS                             & \textbf{0.82} $\pm$ 0.12                      & 1.63 $\pm$ 5.43                      & \textbf{0.86} $\pm$ 0.19                     & \textbf{1.00} $\pm$ 3.00                         \\
2                                 & NVAUTO                            & \textbf{0.82} $\pm$ 0.12                      & 1.63 $\pm$ 4.27                      & 0.80 $\pm$ 0.23                      & 2.00 $\pm$ 2.75                         \\
3                                 & SWAN                              & 0.81 $\pm$ 0.15                      & 1.94 $\pm$ 4.40                     & 0.80 $\pm$ 0.24                      & 2.00 $\pm$ 2.75                          \\
3                                 & PAT                               & \textbf{0.82} $\pm$ 0.15                      & 1.95 $\pm$ 4.71                     & 0.82 $\pm$ 0.30                     & \textbf{1.00} $\pm$ 3.00                         \\
5                                 & CTRL                              & 0.80 $\pm$ 0.14                       & 2.15 $\pm$ 5.32                    & 0.79 $\pm$ 0.23                    & 1.50 $\pm$ 3.00                       \\
6                                 & VICOROB                           & 0.78 $\pm$ 0.18                      & 2.04 $\pm$ 5.29                      & 0.74 $\pm$ 0.31                     & 2.00 $\pm$ 3.00                         \\
7                                 & PLORAS                            & 0.76 $\pm$ 0.16                      & 1.69 $\pm$ 4.71                      & 0.73 $\pm$ 0.33                    & 2.00 $\pm$ 3.00                         \\
8                                 & MIRC                              & 0.76 $\pm$ 0.21                      & 2.78 $\pm$ 6.12                     & 0.67 $\pm$ 0.27                     & 2.00 $\pm$ 4.00                         \\
9                                 & CREATIS                           & 0.71 $\pm$ 0.18                      & 3.30 $\pm$ 7.16                      & 0.67 $\pm$ 0.25                     & 3.00 $\pm$ 5.00                         \\
10                                & Dolphins                          & 0.74 $\pm$ 0.21                      & 2.44 $\pm$ 6.15                     & 0.60 $\pm$ 0.38                      & 3.00 $\pm$ 5.75                         \\
11                                & A55972                            & 0.06 $\pm$ 0.22                      & 10.49 $\pm$ 29.17                    & 0.44 $\pm$ 0.44                     & 4.00 $\pm$ 6.78                         \\
12                                & AICONS                            & 0.4 $\pm$ 0.5                       & 18.86 $\pm$ 31.59                    & 0.26 $\pm$ 0.19                     & 5.00 $\pm$ 6.00  \\       \hline

\end{tabular}
\caption{Final ISLES'22 ranking obtained over the unseen, test-phase data. Median $\pm$ interquartile range values are displayed. *Result obtained in an emulated post-challenge setting by including an ensemble algorithm with the submissions of teams SEALS, NVAUTO and SWAN and by re-computing the challenge ranking. The best median value of each metric is displayed in bold. Statistical significance maps for each metric using Wilcoxon signed-rank tests are available in Supplementary Material S3. DSC: Dice Similarity Coefficient (DSC); F1 score: lesion-wise F1 score; AVD: absolute volume difference; ALD: absolute lesion count difference. }
\label{table:ranking}
\end{center}
\end{table}

\subsection*{Solutions employed by the best-ranked algorithms}
\textbf{Team `SEALS'}.
The participants utilized DWI and ADC images as input for their algorithm. Image pre-processing involved the resampling of the scans to a 1$\times$1$\times$1 mm$^{3}$ voxel resolution, followed by a z-score image normalization. The nn-UNet pipeline \cite{isensee2021nnu} was employed for training a 3D full-resolution UNet. A 1/7 subset of the training dataset was held out to evaluate the performance of the model. The remaining 6/7 parts of the dataset were used to train models through 5-fold cross-validation. A combined Dice loss with categorical cross-entropy was used. Data augmentation transforms were used, including image flipping and Gaussian noise addition. The final submission to the challenge was an ensemble of the five trained models. The algorithm is publicly available at \url{https://github.com/Tabrisrei/ISLES22_SEALS}.

\noindent \textbf{Team `NVAUTO'}.
The team proposed an automated 3D semantic segmentation solution implemented with Auto3DSeg \cite{auto3dseg}. The algorithm automated most deep learning steps and decision choices. DWI and ADC images were used as input to the model after voxel resampling to 1$\times$1$\times$1 mm$^{3}$ and z-score normalization. SegResNet \cite{myronenko20193d} models were trained through 5-fold cross-validation. Several augmentation transforms were used, including flipping, rotation, scaling, smoothing, intensity-scaling and -shifting. Random cropped patches of dimensions 192$\times$192$\times$128 were adopted. The models were trained on an 8-GPU NVIDIA V100 machine with an AdamW optimizer and unitary batch size. Dice loss with focal loss using deep supervision was used as a loss function. The model was first pre-trained on the BRATS 2021 dataset \cite{baid2021rsna}. The final algorithm was an ensemble of 15 models obtained through a 3-time 5-fold cross-validation. The algorithm is publicly available at \url{https://github.com/mahfuzmohammad/isles22}.

\noindent \textbf{Team `SWAN'}.
The participants used the \emph{Factorizer} \cite{ashtari2023factorizer} algorithm to construct an end-to-end, linearly scalable model for stroke lesion segmentation. Factorizer is a family of models that leverage the power of Non-negative Matrix Factorization (NMF) to extract discriminative and meaningful feature maps. The algorithm uses a differentiable NMF layer that can be back-propagated during the training of deep learning models. A \emph{Factorizer} block is constructed by replacing the self-attention layer of a vision transformer block \cite{dosovitskiy2020image} with an NMF-based module and then integrating them into a U-shaped architecture with skip connections. The participants used a Swin Factorizer, which combines NMF with the shifted-window idea inspired by Liu et al. \cite{liu2021swin} to effectively exploit local context. Preprocessing involves FLAIR-to-DWI image registration using Elastix \cite{klein2009elastix} and z-score normalization. Various data augmentation techniques were performed, including random affine transforms, flips, and random intensity scalings. Deep supervision was used at the three highest decoder resolution levels for training the models. The final challenge submission was an ensemble of Swin Factorizers and UNet models with residual blocks \cite{he2016deep} (a.k.a ResU-Net) obtained through 5-fold cross-validation. The pipeline and pre-trained models are available at \url{https://github.com/pashtari/factorizer-isles22}.

Details about the remaining team algorithms are available in Supplementary Material S4.

\subsection*{From challenge to solution: A clinically relevant algorithm}

We aim to test the hypothesis that algorithms devised through a challenge might indeed be relevant for downstream clinical tasks. To do so, we deploy an ensemble algorithm combining technical advantages from different teams' solutions. The hypothesis is tested in two steps. First, we evaluate the ensemble algorithm over diverse clinical and imaging scenarios of the challenge test set to expose potential suboptimal or biased performance toward specific data subgroups. Disease and imaging confounders such as the imaging center, ischemic lesion size, stroke phase, type of stroke pattern/configuration, and vascular territory affected are considered. Second, we test the ensemble algorithm over an independent, large external stroke dataset and assess the lesion segmentation performance and their clinical relevance in real-world data. In the following subsections, we focus on each of these aspects.

\subsubsection*{Can deep learning identify ischemic lesions in scans from an unseen imaging center?}

\begin{figure}[!t]
\begin{center}
\includegraphics[width= 17.17cm]{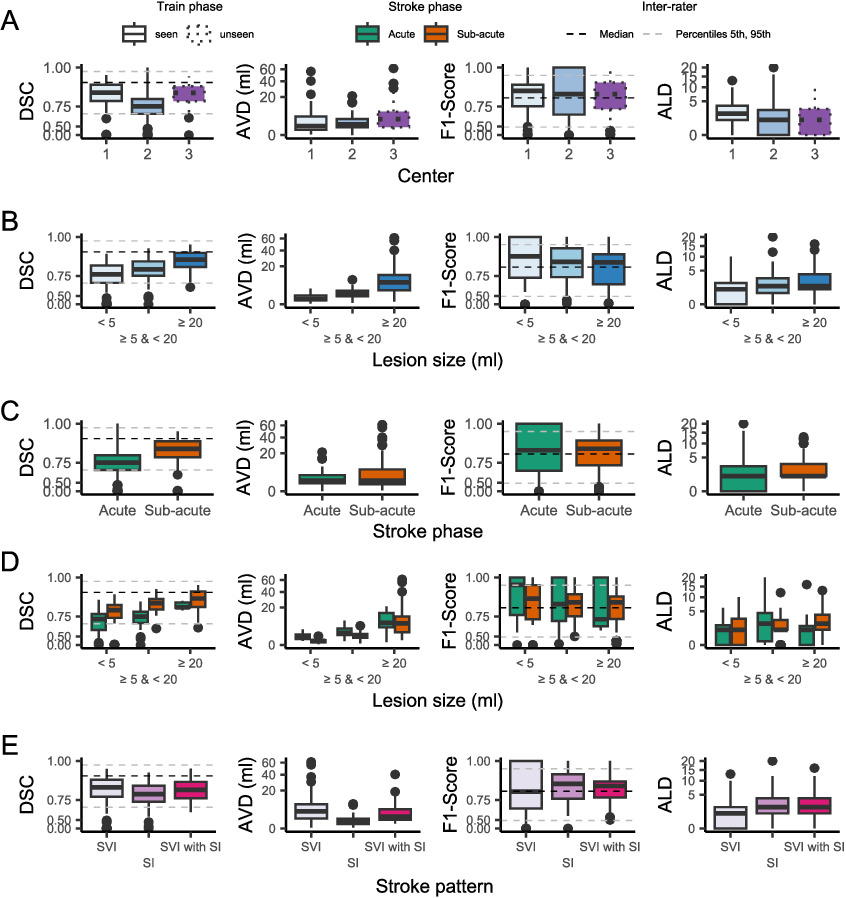}
\caption{Test set performance metrics obtained by the ensemble deep learning algorithm. A) Performance by imaging center. Data is grouped by the center where the images come from ($\#1$, $\#2$, or $\#3$) and by a \emph{seen} or \emph{unseen} label indicating if images from the same center were used for training the models. B) Performance by lesion size. C) Performance by stroke phase (acute or sub-acute).  D) Performance by the stroke phase (acute or sub-acute) grouped by lesion size. E) Performance by stroke pattern subgroups, including single vessel infarcts, scattered infarcts based on micro-occlusions, and single vessel infarcts with accompanying scattered infarcts. 5$\mathrm{^{th}}$, 50$\mathrm{^{th}}$, and 95$\mathrm{^{th}}$ inter-rater variability percentiles are plotted in dashed lines for Dice and F1 score. SVI: single vessel infarct; SI: scattered infarcts based on micro-occlusions; SVI with SI: single vessel infarct with accompanying scattered infarcts. DSC: Dice Similarity Coefficient; F1 score: lesion-wise F1 score; AVD: absolute volume difference; ALD: absolute lesion count difference. y-axes are displayed using a non-linear scale to enhance data visibility.}

\label{fig:generalization}
\end{center}
\end{figure}

Algorithmic robustness to out-of-domain data (unseen during the model's development phase) is crucial for evaluating the algorithm's transferability to real-world centers. Figure \ref{fig:generalization}(A) shows how the ensemble model performs over test-phase data from seen (centers $\#$1 and $\#$2) and unseen (center $\#$3) centers during the development of the algorithm. The distribution of the metrics obtained over the unseen center $\#$3 is similar to the metric's distribution obtained over the seen center $\#$1, suggesting an overall good generalization to new center data (Dice p-value = 0.73, F1 score p-value = 0.60, ALD p-value = 0.42, AVD p-value = 0.08, Wilcoxon rank-sum tests). The performance obtained over the seen center $\#$2 is lower in terms of Dice score compared to the center $\#$1 (p-value < 0.001, Wilcoxon rank-sum test). The F1 score, AVD and ALD metrics are similar between centers $\#$1 and $\#$2 (F1 score p-value = 0.60, ALD p-value = 0.26, AVD p-value = 0.28, Wilcoxon rank-sum tests). The lower Dice scores in center $\#$2 can be explained by two cohort confounders. Firstly, the scans from center $\#$2 include smaller lesion volumes than scans from the other two centers \cite{hernandez2022isles} (p-value = 0.039 for center $\#1$ \emph{vs} center $\#2$, p-value = 0.001 for center $\#2$ \emph{vs} center $\#3$, p-value = 0.56 for center $\#1$ \emph{vs} center $\#3$, Wilcoxon rank-sum tests). Figure S5 (see Supplementary Materials) shows the non-linear, monotonic correlation between lesion size and Dice scores for the test set data. The fact that larger objects (i.e., brain lesions) benefit from higher Dice scores is well known and, therefore, is associated with the found results \cite{maier2022metrics, reinke2023understanding}. Secondly, unlike the train phase data, which considers scans acquired in the sub-acute stroke phase after reperfusion treatment, the test set scans from center $\#$2 are acquired in the acute stroke phase, before the patient's reperfusion treatment, which is known to be a harder task for the algorithms \cite{maier2017isles}. The following sections analyze both of the aforementioned confounding factors.
It is worth noting a potential third confounder related to the imbalance in training data (4:1 for centers \#1:\#2, as shown in Table \ref{table: dataset-summary}). However, we chose to disregard this factor because the model demonstrated strong generalization abilities to scans from an entirely new center. 

\subsubsection*{Can deep learning identify stroke lesions of variable size?}

We test the ensemble model's performance over lesions smaller than 5 ml, lesions larger or equal to 5 ml but smaller than 20 ml, and lesions larger or equal to 20 ml. The performance metrics for the ensemble algorithm over these groups are shown in Figure \ref{fig:generalization}(B). The relationship between lesion size and metrics like Dice, AVD, and ALD is anticipated; larger lesions typically yield higher Dice values, while AVD and ALD tend to increase with lesion size. Despite this, the ensemble algorithm demonstrates strong generalization performance across varying ischemic lesion sizes, achieving comparable ischemia detectability as measured by F1 scores.

\begin{figure}[!b]
\begin{center}
\includegraphics[width=13cm]{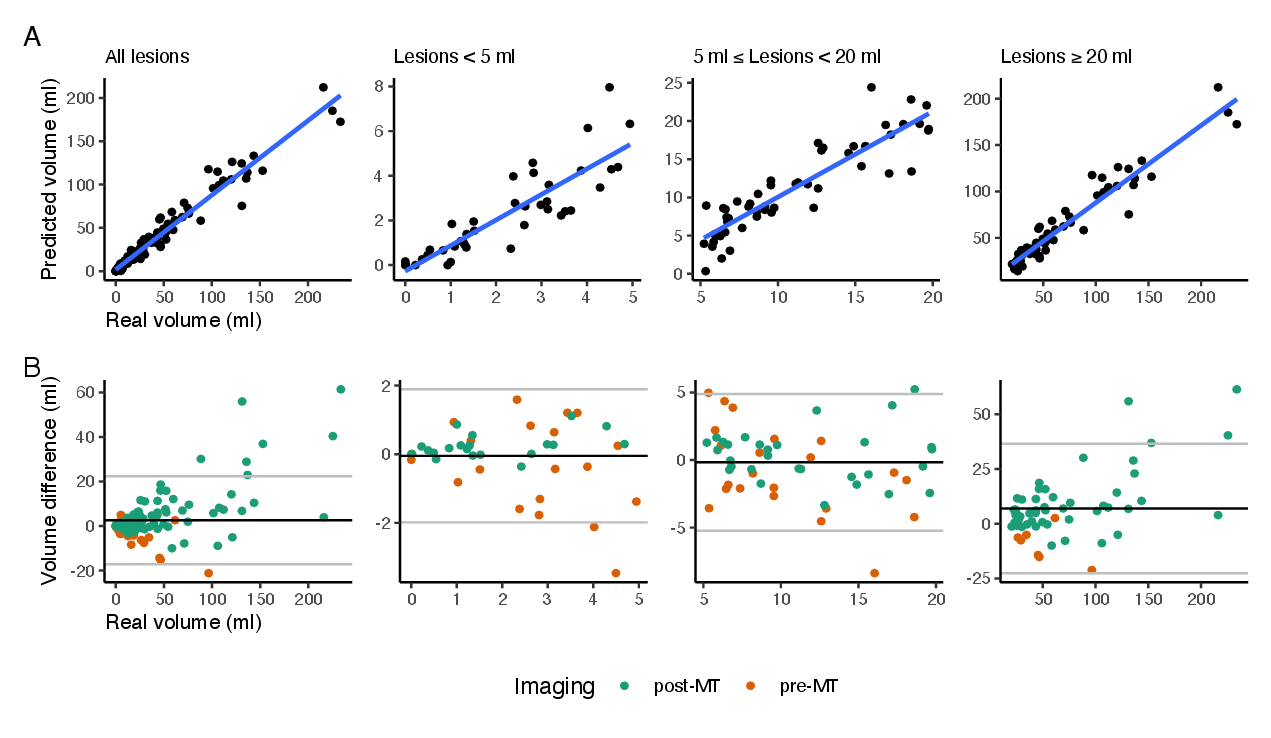}
\caption{Volumetric lesion agreement (deep learning ensemble vs. ground truth) across different ischemic lesion sizes. There is a high volumetric agreement for all lesion sizes, with Pearson $r$ = 0.98 when considering all lesion sizes, $r$ = 0.87 for lesions smaller than 5 ml, $r$ = 0.90 for lesions equal or larger than 5 ml but smaller than 20 ml, and $r$ = 0.96 for lesions larger or equal than 20 ml. A) Scatter plots (predicted volumes vs. ground truth volumes). In blue, a linear regression line is shown. B) Bland-altman plots. The y-axis (volume difference) is calculated as the difference between the real (ground truth) volume and the deep learning (predicted) volume. The black line is plotted at the mean of the volume difference. The gray lines are plotted at $\pm 1.96$ times the standard deviation of the volume difference. MT: mechanical thrombectomy.}

\label{fig:volume_plots}
\end{center}
\end{figure}

Figure \ref{fig:volume_plots} shows the volumetric agreement between the ground truth and the ensemble algorithm predictions for different lesion sizes. There is a high volumetric agreement for all lesion sizes, with Pearson $r$ = 0.98 for the entire test set, $r$ = 0.87 for lesions smaller than 5 ml, $r$ = 0.90 for lesions equal or larger than 5 ml but smaller than 20 ml, and $r$ = 0.96 for lesions larger or equal than 20 ml. The ensemble model thus demonstrated robustness towards diverse stroke lesion sizes.



\subsubsection*{Can deep learning identify ischemia in acute and sub-acute scans?}

In order to assess the generalizability of the algorithm to diverse stroke phases, we split the test set scans into two subgroups: acute (i.e., scans that were acquired as part of the acute stroke diagnostics, within a few hours of stroke onset and before thrombectomy treatment) and sub-acute (i.e., scans acquired within days after stroke onset and after thrombectomy treatment). In Figure \ref{fig:generalization}(C), the performance metrics for the two subgroups are shown. It can be observed that the ensemble model predicts acute scans with similar lesion-wise F1 scores (p-value = 0.45, Wilcoxon rank-sum test) but with lower Dice scores (p-value < 0.001, Wilcoxon rank-sum test) than the sub-acute group. On the contrary, the performance in terms of absolute volume difference and lesion count difference is better for the acute stroke group than for the sub-acute stroke group. These trends are partially due, as earlier introduced, to the lower overall lesion size of the acute group compared to the sub-acute one. However, it remains unclear if the lesion size is the sole responsible for this behavior and what the role of the stroke phase is, especially considering that the training dataset exclusively comprises sub-acute scans. To get insights about it, the test set is grouped considering both the variables: lesion size and the stroke phase. Figure \ref{fig:generalization}(D) shows the corresponding performance metrics. It can be seen that even when splitting the scans using matched lesion-size groups, the lower Dice and AVD performance persists. This indicates that the decline in performance may be attributed to the earlier acute phase of the disease, which was not included in the models' development phase.\\
Bland-Altman plots are shown in Fig. \ref{fig:volume_plots}(B). The percentiles [5th, 50th, 95th] for the volume difference are [-12.2, -1.0, 3.4] ml for the acute group and [-2.6, 1.0, 28.9] ml for the sub-acute one. There is an excellent agreement between the ground truth and predicted lesion volumes for both groups (Pearson $r$ = 0.99 for the acute group and $r$ = 0.98 for the sub-acute one).  

\subsubsection*{Can deep learning predict different stroke clinical patterns?}

\begin{table}[!b]
\begin{center}
\begin{tabular}{ccccc}
                                \multicolumn{1}{l}{}       & \multicolumn{3}{c}{Stroke pattern} & \multicolumn{1}{l}{}        \\ \cline{2-4}
                               &  \makecell{SVI \\(n=62)} & \makecell{SI based on \\ micro-occlusions \\ (n=48)} & \makecell{SVI with\\ accompanying SI \\ (n=38)} & \makecell{All stroke \\(n=148)}   \\ \hline
                               Team & \multicolumn{3}{c}{F1 score} & \multicolumn{1}{c}{Balanced Accuracy}     \\ \cline{2-4}
\multicolumn{1}{c}{SEALS}                     & 87.6                & 75.6                                                          & 68.8                                            & 78.1                \\
\multicolumn{1}{c}{NVAUTO}                   & \textbf{88.1}       & 78.6                                                 & 68.1                                                     & 78.9   \\
\multicolumn{1}{c}{SWAN}                        & 85.0                & 75.9                                                          & 68.2                                                     & 76.2 \\
\hline
\multicolumn{1}{c}{Ensemble}                       &    87.6             &                                                 \textbf{91.8}          &                                               \textbf{81.6}       &   \textbf{86.9}  \\
\hline
\end{tabular}

\begin{tabular}{ccccccc}
\multicolumn{1}{l}{} & \multicolumn{5}{c}{Vascular territory} & \multicolumn{1}{l}{\textbf{}}                \\ \cline{2-6}
\multicolumn{1}{l}{} &  \begin{tabular}[c]{@{}c@{}}MCA\\  (n=97)\end{tabular} & \begin{tabular}[c]{@{}c@{}}PCA \\ (n=23)\end{tabular} & \begin{tabular}[c]{@{}c@{}}ACA\\  (n=4)\end{tabular} & \makecell{Pons / Medula \\ (n=4)}& \makecell{Cerebellum \\ (n=20)} & \begin{tabular}[c]{@{}c@{}}All stroke\\  (n=148)\end{tabular} \\ \hline
Team                 & \multicolumn{5}{c}{F1 score} & \multicolumn{1}{c}{Balanced Accuracy}                                                                                  \\ \cline{2-6}
SEALS                &  97.9                                         & 93.3                                                  & 88.9                                        & 80.0                                                                             & \textbf{97.6}                                                                    & 97.4                                       \\
NVAUTO               &  97.9                                         & \textbf{95.7}                                         & 80.0                                                 & 88.9                                                                             & 97.4                                                                             & 97.3  \\
SWAN           &  96.8                                                  & 93.3                                                  & 66.7                                                 & \textbf{100.0}                                                                   & \textbf{97.6}                                                                    & 92.2                                                  \\ \hline
Ensemble           &      \textbf{98.4}                                               &        93.3                                           &            \textbf{100.0}                                      &                     88.9                                               &            \textbf{97.6}                                                         &       \textbf{97.6} \\ \hline 
 
\end{tabular}
\caption{Algorithmic classification performance of stroke patterns (above) and vascular territories (below). The ensemble algorithm is notably superior than any individual solution in identifying the stroke pattern and the vascular territory. SVI: single vessel infarct; SI: scattered infarcts. All metrics are reported in percentage values. MCA: middle cerebral artery; ACA: anterior cerebral artery; PCA: posterior cerebral artery. All metrics are reported in percentage values.}
 \label{table:stroke_classif_results}
\end{center}
\end{table}

We evaluate whether the ensemble algorithm is affected by diverse stroke lesion patterns. With this aim, the test-phase scans are classified into three stroke sub-groups: single vessel infarcts (SVI), scattered infarcts based on micro-occlusions, and SVI with accompanying scattered infarcts. First, looking for a potential bias towards a specific stroke subgroup, the algorithm performance is evaluated in a subgroup-stratified approach. Second, we frame the problem into a clinically relevant question: Can the ensemble algorithm identify the stroke subgroup?
In Fig. \ref{fig:generalization}(E), the lesion segmentation performance of the ensemble algorithm is shown for each metric and for each type of stroke pattern. A similar performance in terms of Dice score and F1 score for the different stroke subgroups can be appreciated. The lower AVD seen in the group \emph{scattered infarcts based on micro-occlusions} is due to the fact that emboli are typically smaller lesions than SVI and, therefore, this group includes scans with smaller lesion volumes (percentiles [5$\mathrm{^{th}}$, 50$\mathrm{^{th}}$, 95$\mathrm{^{th}}$] of [0.9, 4.4, 36.9] ml) compared to SVI lesions (percentiles [5$\mathrm{^{th}}$, 50$\mathrm{^{th}}$, 95$\mathrm{^{th}}$] of [1.5, 24.9, 137.9] ml) and SVI with scattered infarcts (percentiles [5$\mathrm{^{th}}$, 50$\mathrm{^{th}}$, 95$\mathrm{^{th}}$] of [6.4, 24.5, 134.5] ml). Moreover, the SVI group exhibits lower ALD since their scans have, by definition (see Section \emph{Methods}), less disconnected ischemic lesions.
Next, the ensemble model's capability to predict each scan's stroke subgroup is evaluated. Prediction of the stroke subgroup is generated by applying a heuristic rule defined by radiologists to the stroke masks (details of the classification criteria are available in the \emph{Methods} section). Results for each of the top-3 ranked methods, as well as for the ensemble algorithm, are summarized in Table \ref{table:stroke_classif_results}. The most challenging scans to identify are the ones exhibiting an SVI with accompanying scattered infarcts. It is also worth noting that the solution submitted by the team `NVAUTO' (which ranked second in the challenge) yields a better stroke pattern classification performance than the other challenge submissions. The best overall performance is obtained by the ensemble algorithm, which remarkably outperforms any single challenge algorithms (balanced accuracy of 86.9\% for the ensemble method compared to the 78.9\% achieved by the team `NVAUTO'), thus demonstrating a strong capability to classify stroke sub-groups.

\subsubsection*{Can deep learning identify the ischemic vascular territory?}
 In this experiment, we evaluate whether the algorithms can identify the affected vascular territory among the middle, anterior, and posterior cerebral arteries, the vasculature of the cerebellum, and the vasculature of the pons/medulla. To this end, we quantify through the predicted lesion masks the lesion load per vascular territory from a reference atlas of vascular territories. Then, the territory with the absolute largest lesion volume is considered the most affected territory and is compared with the vascular territory affected in the ground truth masks. In Table \ref{table:stroke_classif_results}, the results from this experiment are shown. Overall, the challenge algorithms accurately predict most vascular territories. The solution submitted by the winner of the challenge (team `SEALS') obtains the best performance in this task when compared to the other teams. The best overall performance is obtained, again, with the ensemble algorithm, yielding a remarkable 97.6\% of balanced accuracy and F1 score $>$ 88\% for each considered vascular territory. Our results show that the ensemble algorithm can accurately identify the impacted vascular territory.

\begin{figure}[!b]
\begin{center}
\includegraphics[width=17.15cm]{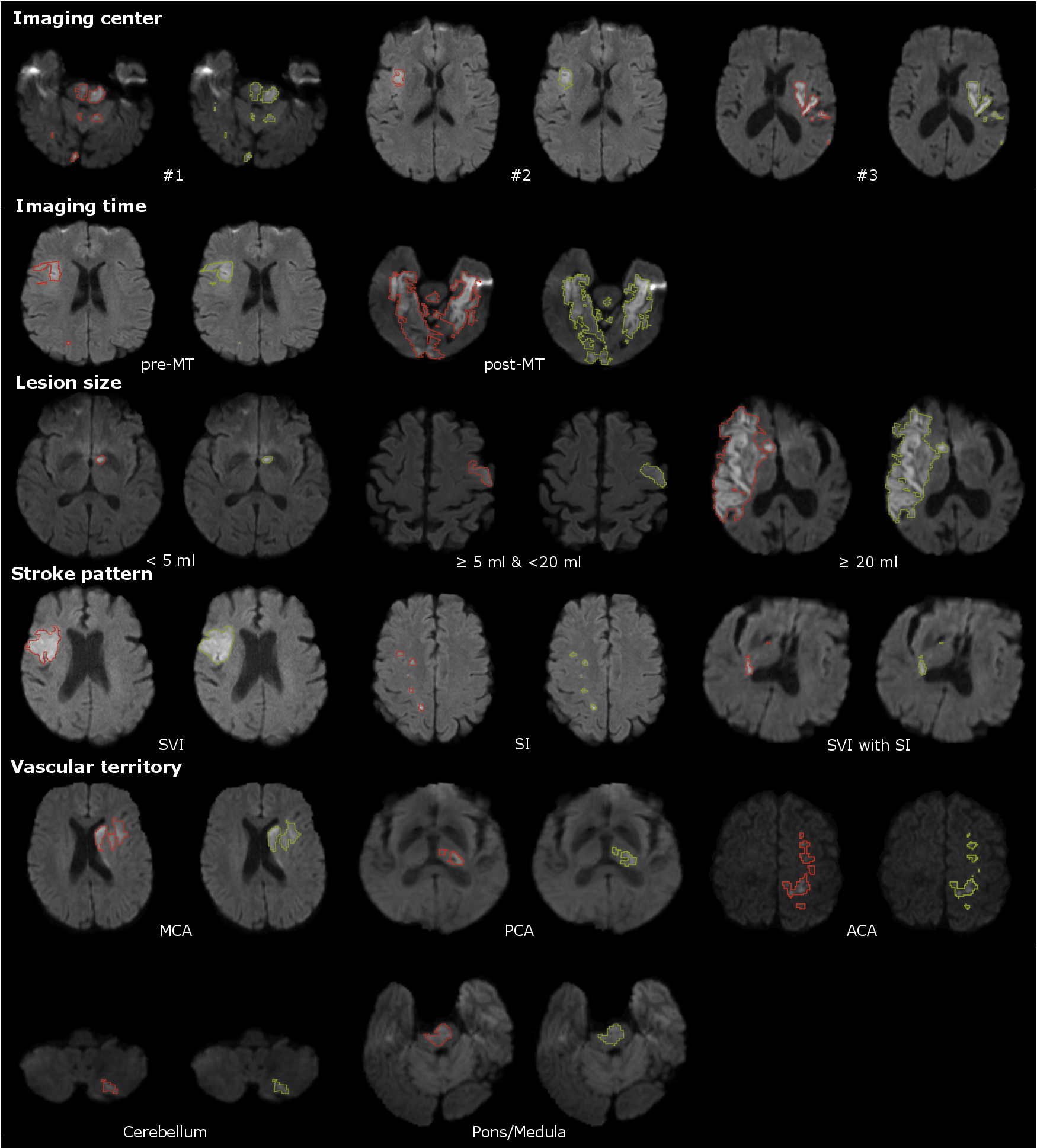}
\caption{Lesion ground truth (red) and predictions (green) obtained with the ensemble algorithm over unseen test-phase scans. The developed algorithm closely aligns with expert annotations across various types of stroke patterns and configurations. Results are grouped by healthcare center, imaging time, lesion size, stroke pattern, and vascular territory affected. The visualized images are scans with median Dice performance (or with the closer Dice to the median value) for the different groups. MT: mechanical thrombectomy; SVI: single vessel infarct; SI: scattered infarcts based on micro-occlusions. SVI with SI: single vessel infarct with accompanying scattered infarcts. MCA: middle cerebral artery; ACA: anterior cerebral artery; PCA: posterior cerebral artery.}
\label{fig:qualitative}
\end{center}
\end{figure}

\subsection*{Framing the findings: comparison with inter-rater performance}
Two expert neuroradiologists annotated ten randomly sampled scans from the ISLES'22 training set \cite{hernandez2022isles}. When comparing their delineations against the ground truth masks, they achieved a median $\pm$ interquartile range Dice score of 0.92 $\pm$ 0.16 and a lesion-wise F1 score of 0.82 $\pm$ 0.30. The ensemble algorithm yielded, over the entire test set, a Dice score of 0.82 $\pm$ 0.12 and an F1 score of 0.86 $\pm$ 0.21. 
Figure \ref{fig:qualitative} shows the median Dice MRI scan (or the closest to the median) for diverse stroke scenarios. The visual results of the predicted ischemic masks suggest that the ensemble predictions (green delineations) closely follow the manually segmented lesions (red delineations), exposing the algorithm's capability to cope with diverse types of stroke patterns and configurations. The quantitative and qualitative results suggest robustness towards heterogeneous clinical and imaging scenarios.

\subsection*{Qualitative results: Deep learning versus experts in a Turing-like test}
\begin{figure}[!t]
\begin{center}
\includegraphics[width = 12cm]{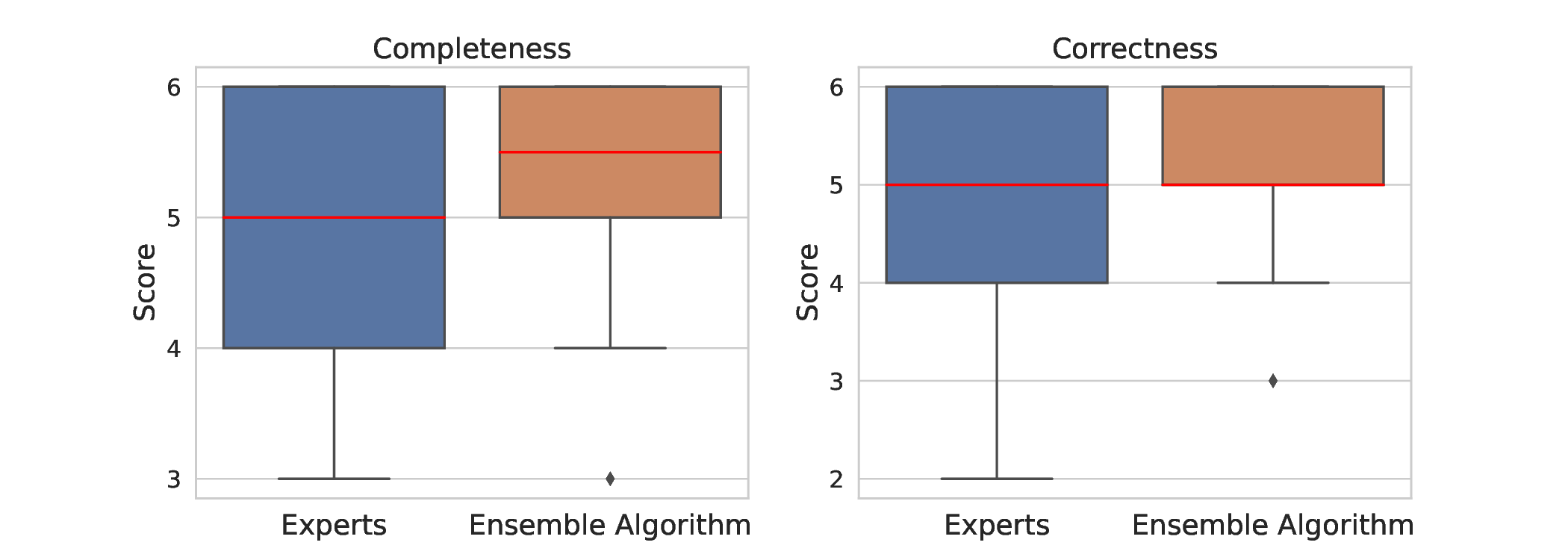}
\caption{Qualitative lesion segmentation results obtained in a Turing-like test. Neuroradiologists prefer lesions delineated by the ensemble algorithm over manual expert delineations. Score values range between 1 and 6 (worst and best quality scenarios, respectively).}
\label{fig:turing}
\end{center}
\end{figure}

In a Turing-like test, nine experienced radiologists rated the stroke segmentation quality of the ISLES'22 test data. Radiologists received forty or forty-one randomized images with each image being delineated either by an expert or by the ensemble algorithm and were asked to rate the \emph{completeness} and \emph{correctness} of the lesion masks in a 1-to-6 (worst-to-best) scale. Boxplots of this experiment are shown in Fig. \ref{fig:turing}. Interestingly, the ensemble algorithm exhibits statistically significantly higher ratings than the experts (p-value = 0.02 when considering the segmentation \emph{completeness} and p-value <  0.001 when considering the segmentation \emph{correctness}, Wilcoxon signed-rank tests). The observation that experts find deep learning segmentations to be qualitatively superior to manually traced ones is unsurprising, given that similar findings were reported in prior research \cite{kofler2021we}. 

\subsection*{External validation and clinical relevance}
We tested the ISLES'22 ensemble algorithm over the largest available acute and early sub-acute ischemic stroke cohort (N = 1686) retrieved from the Johns Hopkins Comprehensive Stroke Center \cite{liu2023large}. The data were collected over ten years utilizing eleven magnetic field scanners operating at either 1.5T or 3T, sourced from four different vendors. Due to continual technological advancements over the last decade, the dataset comprises diverse MRI acquisition protocols and machine technologies, leading to strong variations in image quality (such as resolution, signal-to-noise ratio, and contrast-to-noise ratio). This makes this dataset uniquely suited to evaluate the generalizability of algorithms. Ischemic lesion masks were annotated by the dataset authors through manual delineation. Patient age was 62.5 $\pm$ 13.3 years. 907 patients (53.8\%) were male. Reported race or ethnicity included 753 (44.7\%) Black or African American, 490 (29.1\%) White, and 40 (2.4\%) Asian. 876 (52.0\%) MRI scans were acquired after thrombolysis treatment with intravenous tissue-type plasminogen activator. 

Table \ref{table:jhcsc_results} summarizes the algorithmic results in this challenging external test set. The performance of the individual algorithms well reflects the patterns observed in the ISLES'22 test set (Table \ref{table:ranking}): While the team 'SEALS' leads the lesion-wise detection in terms of F1 score, the team `NVAUTO' leads the segmentation performance in terms of Dice scores. Importantly, the ensemble algorithm consistently outperforms all individual algorithms, exhibiting statistical significance across all metrics. Furthermore, the ensemble algorithm exhibits enhanced robustness and effectively mitigates the weaknesses observed in the single algorithms. It surpasses the `SEALS' algorithm in segmentation performance, achieving 8\% and 1\% higher 5th and 50th Dice percentiles, respectively, and it outperforms both the `NVAUTO' and `SEALS' methods in lesion-wise detection, with a 6\% higher 50th percentile F1 score than `NVAUTO' and a 2\% higher 50th percentile F1 score than `SEALS'. Overall, the performance achieved on the external Johns Hopkins dataset closely aligns with the results obtained on the ISLES'22 test set. Specifically, the median $\pm$ interquartile range Dice scores and lesion detection F1 scores are 0.82 $\pm$ 0.15 and 0.86 $\pm$ 0.33, respectively, as outlined in Table \ref{table:jhcsc_results}. There are no statistically significant differences in segmentation performance between the external dataset results and the ISLES'22 ones (Dice coefficient p-value = 0.46, F1 score p-value = 0.66, Wilcoxon rank-sum tests).
The correlation between the lesion volumes estimated with the ensemble algorithm and those manually obtained by experts is very high (Pearson $r$ = 0.97). Moreover, we evaluated the association between the estimated lesion volumes of the ensemble algorithm with the National Institutes of Health Stroke Scale (NIHSS) at patient admission (N = 999) and with the modified ranking scales (mRS) at 90-day follow-up (N = 782). We observe a slightly higher correlation between volumes and clinical scores when using the ensemble algorithm-predicted masks (NIHSS $r$ = 0.55 and 90-day mRS $r$ = 0.41, Pearson correlation coefficients) compared to the manually delineated lesion ones (NIHSS $r$ = 0.54 and 90-day mRS $r$ = 0.39, Pearson correlation). These findings show that the proposed algorithm can derive downstream clinical scores at a level comparable to or potentially surpassing those derived by radiologists.

\begin{table}[!t]
\begin{center}
\begin{tabular}{ccccccccc}
\hline 
\textbf{Algorithm} & \textbf{DSC $\uparrow$} & \textbf{p-value}& \textbf{F1 $\uparrow$}& \textbf{p-value} & \textbf{AVD (ml) $\downarrow$} & \textbf{p-value}& \textbf{ALD $\downarrow$} & \textbf{p-value} \\
\hline 
\multirow{2}{*}{Ensemble} & \textbf{0.82} $\pm$ 0.15 & \multirow{2}{*}{-} & \textbf{0.86} $\pm$ 0.33 & \multirow{2}{*}{-} & 0.86 $\pm$ 3.96 & \multirow{2}{*}{-} & \textbf{1.00} $\pm$ 2.00 \\
         & [0.45, 0.94]         && [0.4, 1.00]         && [0.03, 18.57]  && [0.00, 10.00] \\
\hline 
\multirow{2}{*}{SEALS}    & 0.81 $\pm$ 0.16          &\multirow{2}{*}{< 0.001}& 0.84 $\pm$ 0.33         &\multirow{2}{*}{< 0.001} &0.93 $\pm$ 3.98    & \multirow{2}{*}{< 0.001}& \textbf{1.00} $\pm$ 2.00 & < 0.001 \\
         & [0.37, 0.94]         & & [0.4, 1.00]         & &[0.03, 19.37]  & & [0.00, 9.00] \\
\hline
\multirow{2}{*}{NVAUTO}   & \textbf{0.82} $\pm$ 0.15 &\multirow{2}{*}{< 0.001} & 0.80 $\pm$ 0.33   &\multirow{2}{*}{< 0.001} & \textbf{0.84} $\pm$ 3.87 &\multirow{2}{*}{0.038} & \textbf{1.00} $\pm$ 3.00 &\multirow{2}{*}{< 0.001} \\
          &[0.47, 0.94]         & &[0.4, 1.00]         & &[0.03, 18.50]  & & [0.00, 10.00] \\
\hline
\multirow{2}{*}{SWAN}     & 0.79 $\pm$ 0.20    & \multirow{2}{*}{< 0.001}& 0.80 $\pm$ 0.33&  \multirow{2}{*}{< 0.001}& 1.01 $\pm$ 4.11    & \multirow{2}{*}{< 0.001}& \textbf{1.00} $\pm$ 3.00 &\multirow{2}{*}{< 0.001} \\
         & [0.10, 0.92]   & & [0.29, 1.00  & & [0.03, 20.65]  & & [0.00, 11.00]  \\
\hline

\end{tabular}
\caption{Algorithmic performance in the Johns Hopkins Comprehensive Stroke Center dataset. The ensemble algorithm surpasses individual methods when applied to an external, real-world dataset and effectively addresses the lesion segmentation (detection) weaknesses observed in algorithm `SEALS' (`NVAUTO'). Median $\pm$ interquartile range and [5th and 95th percentile] values are shown. DSC: Dice Similarity Coefficient (DSC); F1 score: lesion-wise F1 score; AVD: absolute volume difference; ALD: absolute lesion count difference. The best median values are shown in bold. We conducted two-sided Wilcoxon's signed rank tests with subsequent false discovery rate adjustment using the Benjamini-Hochberg method to compare metrics.}
\label{table:jhcsc_results}
\end{center}
\end{table}

\section*{Discussion}
Accurate segmentation of ischemic stroke lesions from brain MRI is crucial for timely diagnosis, treatment planning, and patient follow-up. Deep learning offers a promising avenue to support radiologists by enabling faster, more objective, and potentially more accurate MRI analysis. This study addresses this challenge by proposing a clinically meaningful and generalizable deep learning algorithm for ischemia segmentation. To foster development and rigorously assess candidate solutions, we organized the international ISLES'22 medical segmentation challenge. ISLES'22 served as a powerful platform for rapid algorithm benchmarking and identification of promising approaches, including the one presented here.

The following discussion delves into two key aspects. First, we explore how ISLES'22 facilitated identifying strong AI-based methods. Second, we discuss the real-world readiness of the ensemble algorithm, highlighting its potential for robust performance under variable data conditions.

\subsection*{ISLES'22 Challenge: Benchmarking Stroke Segmentation in a Real-World Setting}
This paper benchmarks ischemic stroke segmentation algorithms developed for the ISLES'22 challenge. The challenge design offers significant advancements over prior iterations by incorporating a large, multi-centric dataset with over 6 times more scans than ISLES'15. This data reflects the real-world heterogeneity of stroke lesions, promoting generalizability. Notably, minimal preprocessing was applied, focusing solely on patient de-identification. This challenged participants to develop end-to-end solutions encompassing all necessary processing steps (e.g., modality selection, registration, normalization), mimicking real-world clinical workflows. This, in turn, discouraged the convergence towards a single, potentially overfitted solution, as can occur with highly curated datasets.
Furthermore, the challenge fostered robust evaluation by employing hidden data for testing. Participant models were presented with unseen MRI scans, preventing both model overfitting and intentional calibration towards specific images.

Choosing appropriate evaluation metrics can be difficult in challenge design. ISLES'22 addressed this by incorporating expert recommendations \cite{maier2022metrics, reinke2023understanding}. As suggested, we employed a balanced mix of technical metrics (e.g., Dice scores) and clinically relevant metrics (e.g., number and volume of predicted ischemic lesions). This comprehensive approach allows for a broader assessment of solution performance and its readiness for real-world clinical applications, such as AI-assisted radiological reading.

The ISLES'22 challenge yielded fascinating insights into the landscape of stroke segmentation algorithms. Interestingly, the challenge leaderboard revealed that even algorithms based on similar CNN architectures and optimization strategies can exhibit variable performance. This reinforces the notion that factors beyond architecture, like hyper-parameter tuning, stochastic optimization, and training data sub-splitting (as in cross-validation), all contribute to model variability, even with a consistent dataset like ISLES'22 \cite{haibe2020transparency}.
However, the challenge also showcased the effectiveness of diverse algorithmic approaches. While achieving similar performance on most metrics, the top three ranked solutions employed different methodologies. The leading two teams utilized distinct CNN architectures (nnUnet \cite{isensee2021nnu} and SegResNet\cite{myronenko20193d}) and loss functions (Dice with binary cross-entropy vs. Dice with focal loss). Notably, the third-ranked solution adopted a completely different approach based on non-negative matrix factorization \cite{ashtari2023factorizer}. This solution also leveraged the FLAIR modality (discarded by the top two teams), necessitating additional FLAIR-to-DWI co-registration.
This remarkable diversity in successful solutions highlights the power of challenges in fostering innovation and the potential of ensemble methods that combine these strengths.

\subsection*{Beyond ISLES'22: Towards automatic ischemic stroke segmentation in the clinical setting}
Our benchmarking process identified promising algorithms that could potentially be deployed in clinical settings. However, to ensure the development of truly reliable AI solutions, a deeper understanding of these algorithms' strengths and limitations is paramount. We addressed this need by extending our analysis beyond simple benchmarking. We conducted a detailed evaluation of the algorithm outputs across various axes of generalization, including $i)$ imaging center, $ii)$ ischemic lesion size, $iii)$ stroke phase, $iv)$ type of stroke pattern/configuration, and $v)$ anatomical location of the ischemia.

Our ensemble algorithm demonstrates robust performance in handling a wide range of image and disease variations. This is evident from the successful generalization to unseen data from a new center, which achieved results similar to those of the trained center. Interestingly, while centers $\#$1 and $\#$3 (seen and unseen, respectively) showed similar metric distributions, a significant difference in Dice scores arose between centers $\#$1 and $\#$2 (both seen during training). This discrepancy can be attributed to two key factors. First, center $\#$2's scans had considerably smaller stroke lesions. Second, these scans were acquired during the acute stroke phase.
This observation highlights a crucial novel finding: the timing of brain imaging relative to stroke onset significantly impacts deep learning model performance. Lesion volumes from acute stroke scans were slightly overestimated compared to sub-acute ones. Despite this difference in Dice scores, the volumetric agreement between predictions and ground truth remained exceptionally high for both groups (Pearson's r = 0.99 and 0.98 for acute and sub-acute, respectively). Ischemia detectability, as measured by lesion-wise F1 scores, also remained statistically similar between groups. These results lead to two key conclusions:
$i)$ The model demonstrates remarkable generalizability across different stroke phases (acute and sub-acute) and imaging centers. This suggests the successful capture of stroke lesion variability, avoiding reliance on center-specific MRI features.
$ii)$ The stroke phase at scan acquisition demonstrably influences the model. This is understandable as early (acute) scans exhibit different MR characteristics compared to later ones (sub-acute) due to evolving tissue changes. This aligns with established knowledge about how DWI and ADC values fluctuate with stroke progression \cite{fung2011mr}. Similarly, DWI sensitivity (specificity) ranges between 73\% (92\%) 3 hours from the stroke event to 92\% (97\%) 12 hours from the stroke event \cite{chalela2007magnetic, merino2010imaging}. Furthermore, false negatives may also increase with early DWI acquisition \cite{sylaja2008expect, merino2010imaging}.

Volumetric analysis revealed a very high correlation between the predicted lesions and ground truth (Pearson r = 0.98). This agreement remained strong for lesions of varying sizes (Pearson r > 0.85), particularly for larger ones. Notably, the algorithm performed consistently across lesion sizes (similar F1 scores), suggesting reliable detection of both small (e.g., emboli) and large (e.g., large vessel occlusions) strokes.

Beyond lesion volume, our model sheds light on stroke etiology. Traditionally, stroke type and affected vascular territories are crucial for determining the underlying cause, impacting treatment decisions and prevention strategies (e.g., Merino et al.\cite{merino2010imaging}, Kim et al.\cite{kim2014magnetic}). Existing research establishes associations between DWI lesion patterns and stroke causes (e.g. Kang et al.\cite{kang2003association}). For instance, scattered infarcts across multiple territories, single cortico-subcortical lesions, or multiple lesions in the anterior and posterior circulation often indicate cardioembolism \cite{cho2007mechanism, kim2014magnetic, merino2010imaging, kang2003association}. Conversely, large artery atherosclerosis typically presents with lesions in a single vascular territory \cite{merino2010imaging, kang2003association}.
In this context, our deep learning algorithm stands out by accurately characterizing DWI images from a multi-faceted perspective. The model effectively segments ischemia in different stroke sub-groups (SVI, scattered infarcts, and mixed SVI/scattered) with consistent performance. Importantly, the model tackles even small ischemic volumes (e.g., embolic showers) with high detectability (lesion-wise F1 scores comparable to larger lesions). Furthermore, it accurately identifies stroke subgroups and affected vascular territories using predicted lesion masks (multi-class balanced accuracies of $\sim$ 87\% and $\sim$ 98\%, respectively, Table \ref{table:stroke_classif_results}). These findings suggest that the algorithm's outputs extend beyond lesion volume, providing valuable clinical insights into stroke type and underlying cause, ultimately aiding downstream clinical assessments.

To further assess segmentation quality, we employed a blinded evaluation similar to a Turing test. Nine experienced neuroradiologists from various institutions judged the \emph{completeness} (lesion identification) and \emph{correctness} (delineation accuracy) of both manual and algorithmic segmentations. Their preference, aligning with Kofler et al.'s findings for brain tumors \cite{kofler2021we}, consistently favored the ensemble algorithm's outputs. This finding suggests that deep learning algorithms can match or even surpass human experts in identifying brain infarcts on MRI scans.

The key innovation lies in the ensemble approach. Validation on a large, unseen, and highly heterogeneous patient cohort revealed that the ensemble method overcomes limitations of individual algorithms, leading to increased robustness in ischemia detection and segmentation. Notably, the ensemble leverages the strengths of each constituent algorithm, such as NVAUTO's high Dice score and SEALS' high F1 score, to generate segmentations that significantly surpass individual performances. Furthermore, the algorithm's results closely matched those obtained from the ISLES'22 test set, validating the effectiveness of both the challenge dataset and its organization.
These results translate into two significant conclusions. First, the ensemble algorithm demonstrates robustness towards unseen data from a new center, encompassing the vast variability in image and patient characteristics characteristic of real-world settings. This translates to excellent segmentation generalization to out-of-domain data. Second, our analysis demonstrates agreement between lesion volumes obtained from the algorithm and those delineated by experts. Moreover, the algorithm-derived volumes explain clinical scores (admission NIHSS and 90-day mRS) at least as effectively, and potentially even better, than those obtained through manual expert delineation. This implies a significant contribution: we present an ensemble algorithm, derived from a limited-data challenge, that delivers clinically relevant stroke segmentations on real-world patient data. This extends its applicability beyond the challenge setting and demonstrates its value for downstream clinical analysis. We have made the ensemble model publicly available on \url{https://github.com/Tabrisrei/ISLES22_Ensemble} to facilitate broader use.

\subsection*{Limitations and outlook} 
Our work presents a novel ensemble algorithm for stroke lesion segmentation with promising results. However, to ensure generalizability and real-world applicability, further validation is necessary.
We systematically introduced variability to the data to assess the algorithm's performance across different patient populations and stroke scenarios. While we focused on major sources of variation, limitations remain. First, the external testing dataset lacked multi-center representation. Second, the ISLES'22 data consisted solely of European cohorts, while the external dataset, although offering more patient race variability, was limited to the US population.
Addressing these limitations, we believe the ensemble algorithm would significantly benefit from validation on external cohorts collected from diverse medical centers worldwide. Furthermore, inclusion of under-represented patient groups is crucial. We encourage clinical research groups to participate in validating and refining our model to enhance its generalizability and clinical impact.

\section*{Methods}
In order to devise an automatic algorithm that can identify brain ischemia under heterogeneous \emph{real-world} imaging scenarios, the public challenge ISLES'22 is organized. The ISLES'22 challenge enables a fast and extended benchmarking of algorithmic strategies for tackling the task. Therefore, this section is organized in two parts. The first section explains how the challenge is structured. The BIAS (Biomedical Image Analysis ChallengeS) \cite{maier2020bias} guideline is followed. The second section details how the identified algorithms are evaluated to come up with a clinically useful solution.

\subsection*{The ISLES challenge}
\subsubsection*{The ISLES mission}
The ISLES challenge (\url{https://www.isles-challenge.org/}) is a multi-institution, non-profit initiative involving leading neurointerventionalists, radiologists, and researchers in the field of ischemic stroke and medical imaging. From 2015, and with few exceptions, a yearly data challenge has been opened to the research community with the aim of improving accuracy, fairness, and reproducibility of medical imaging algorithms used for the evaluation of ischemic stroke imaging data. 

\subsubsection*{The ISLES'22 task}
 For the 2022 challenge edition, participants are requested to develop fully automatic algorithmic solutions that segment ischemic lesions in hyperacute, acute and early subacute stroke MRI (DWI, ADC, and FLAIR). Algorithms generate, as output, a binary stroke segmentation mask.

\subsubsection*{Challenge organization}
The aims, structure, and organization of the challenge are available for the teams several weeks before the dataset is released. A detailed description of the challenge organization is available in de la Rosa et al.\cite{islesdocument}. The challenge is organized in three phases: a \emph{train}, a \emph{sanity-check}, and a \emph{test} phase. In the train phase, participating teams have six weeks to develop a solution to the task using a labeled MRI dataset. All teams have access to the data at the same time. There are no technical constraints on the employed algorithmic solution. In the sanity-check phase, participants can test their devised Docker solutions over a few train set scans. The aim of this phase is to guarantee that each team's algorithm properly works in the challenge organizers' servers. Therefore, multiple submissions of the algorithms are allowed. In the test phase, participating teams are requested to submit a Docker containing their final algorithmic solution. Teams can submit just a single time to this phase. The algorithm is later run by the challenge organizers over the hidden test set. Algorithmic performance is measured by computing relevant metrics (below detailed) using the predicted segmentations and the ground truth masks. Later, teams are ranked based on their yield performance metrics.

\subsubsection*{Dataset}
The dataset used in this study is customarily devised for the purposes of the challenge. It contains multi-center and multi-scanner data, and it consists of MRI scans (n=400) acquired during the early/late acute and the early sub-acute stroke phase of patients across three European health centers. The train (n=250) and test (n=150) sets include scans from two and three healthcare centers, respectively. As the challenge was focused on post-operative stroke images, the inclusion criteria consisted on patients who underwent MRI imaging after receiving mechanical thrombectomy treatment. Furthermore, as a proof of concept, we aim to test the generalization capability of the devised algorithms over a small, single-center subset of hyper-acute/early acute stroke scans (12.5\%, n=50) before intervention, which represents a third part of the test set. Table \ref{table: dataset-summary} provides a summary of the ISLES'22 dataset. All cases include DWI (b-value=1000 s/mm$^2$), FLAIR, and ADC MRI. The ischemic stroke segmentation ground truths are obtained using an algorithm-human hybrid iterative method and are later revised and refined by one out of three experienced neuroradiologists with more than 10 years of experience (RW, JSK and BW). The MRI images are released in their native acquisition space (ground truth masks are released in the DWI/ADC space) after minimal pre-processing. Thus, pre-processing is solely performed with the purpose of patient de-identification and therefore consists of MRI skull-stripping. The reason for releasing the dataset `as raw as possible' is to encourage the development of algorithms that could deal with real-world raw imaging data, which suffers from a large variability (signal-to-noise, resolutions, variable parameter settings, etc.) and therefore has its own technical limitations (e.g., different MR modalities, as FLAIR and DWI, are not co-registered). In this sense, participants are also challenged to devise end-to-end algorithms that can deal with the pre-processing and curation of the images. For further details about this dataset please check the corresponding data descriptor \cite{hernandez2022isles}.

\begin{table}[]
\begin{center}
    
\begin{tabular}{cccccccc}
Dataset &Phase & \multicolumn{1}{l}{\# Scans} &\begin{tabular}{c} \# Scans by center \\(1 / 2 / 3) \end{tabular}& RT & \begin{tabular}{c}\# Scans \\pre-RT \end{tabular}&\begin{tabular}{c}\# Scans \\post-RT \end{tabular}& Stroke phase\\
\hline
\multirow{2}{*}{ISLES'22\cite{hernandez2022isles}}&Train & 250  &                         200 / 50 / 0             & MT & 0                                   & 250 & Sub-acute                                  \\
& Test  & 150                          & 50 / 50 / 50            &MT  & 50                                  & 100 & Acute and sub-acute                         \\
\hline
JHCSC\cite{liu2023large}& Test  & 1686                          & -            &ivtPA  & 809                                 & 870        & Acute and sub-acute                 \\
\hline

\end{tabular}
\end{center}
\caption{Data summary. The ISLES'22 dataset was used in the challenge competition, while the JHCSC was used as an external testing dataset in a post-challenge setting. JHCSC: Johns Hopkins Comprehensive Stroke Center; RT: reperfusion treatment; MT: mechanical thrombectomy; ivtPA: intravenous tissue plasminogen activator. Further details about the datasets are available in the corresponding data descriptors \cite{hernandez2022isles,liu2023large}.}
\label{table: dataset-summary}
\end{table}

\subsubsection*{Algorithms' evaluation}
\paragraph*{Metrics}
\label{section:metrics}
The algorithmic results are evaluated by comparing the predicted lesions masks with the (manually traced) ground truth masks. Selection of the appropriate metrics to assess the method's performance was heavily discussed within the challenge organization team based on current recommendation guidelines \cite{maier2022metrics, reinke2023understanding}. The aim was to conduct a clinically relevant evaluation unbiased by diverse lesion characteristics and lesion patterns. The following metrics that are well known in the literature for both the research and radiological communities were included: Dice score \cite{dice1945measures} (DSC), the absolute volume difference ($\mathrm{AVD = |Volume_{\,predicted} - Volume_{\,ground truth}|}$), lesion-wise F1 score (defined as in Section \emph{Statistical analysis} by considering instance lesions), and the absolute lesion count difference ($\mathrm{ALD = |\#\, Lesions_{\,predicted} - \#\, Lesions_{\,ground\: truth}|}$). Lesion-wise metrics were computed after isolating disconnected ischemias in the binary masks through connected-component analysis. Note that including complementary segmentation and detection metrics is beneficial and helps evaluate models in a robust, subject, and lesion-wise unbiased scheme \cite{sudre2022valdo}. Dice was used to evaluate semantic segmentation at the entire subject level. Given that segmentation metrics (as DSC) do not provide insights about the clinical utility of the predictions, we also included a general clinical metric of interest as the AVD. Furthermore, lesion detection metrics (which are computed for each instance lesion) were considered since the dataset has scans with multiple and small emboli, which might not drive important changes in the DSC and AVD metrics under the presence of large disconnected infarcts. For example, algorithms missing a small lesion can achieve a very high DSC if a much larger infarct within the same scan is properly segmented. However, an algorithm detecting both lesions (which is clinically desired) might still perform worse in terms of DSC and AVD. Consequently, to penalize algorithms that under- or over-estimate the number of predicted lesions, we evaluate the algorithm's capability to detect each (single) disconnected ischemia through the lesion-wise F1 score and ALD. Implementation details of the four metrics can be found in the challenge Python notebooks\cite{islesrepo, atlasrepo}.

\paragraph*{Ranking}
\label{section:ranking}
The final competition ranking is obtained from the \textit{test} phase, which is \emph{blind} (there is no access for the teams to the MRI images to be predicted) and \emph{single-shoot} (one submission per team is solely allowed), thus completely precluding participants from any sort of overfitting strategy. As similarly done in previous ISLES editions\cite{maier2017isles, winzeck2018isles, hakim2021predicting}, the ranking is performed in a `rank then aggregate' fashion  \cite{maier2018rankings}. In a nutshell, the ranking scheme is obtained as the average ranking position obtained from all the individual rankings by case and by metric. A thousand bootstraps are conducted by repeatedly drawing samples with replacements and recalculating the rankings in each iteration. Furthermore, as a complementary analysis, we calculated the challenge ranking through a thousand bootstraps `aggregate then rank' scheme.

\subsection*{Towards real-world clinical performance}
\subsubsection*{Ensemble algorithm}
We select the top-3 ranked solutions of the challenge and generate new predictions through their ensemble. The predictions are obtained through majority voting. Thus, for each voxel of an MRI scan, the predicted output (i.e., lesion or no lesion) is obtained by the agreed result predicted by at least two of the methods. The algorithms considered in the ensemble were those leading the leaderboard but using different solutions. Hence, to prevent redundant predictions arising from similar solutions (with equivalent architecture and training strategy) submitted by different teams, only the highest-performing implementation is considered for that specific algorithm type.

\subsubsection*{Jeopardizing the solution: Which (and how) real-world variables impact it?}
\label{section:variables}

With the aim of understanding the potential clinical utility of the deep learning solution,  we evaluate whether the method can detect ischemia under diverse disease and imaging scenarios, thus providing insights about its robustness and generalization capability when dealing with diverse ischemic stroke events. With this aim, the test-phase predictions of the ensemble algorithm are evaluated over $i)$ scans coming from an external healthcare center, unseen during model development, $ii)$ scans with diverse ischemic lesion size, $iii)$ scans with ischemia located in diverse vascular territories of the brain, $iv)$ scans with diverse lesion configurations and patterns, and $v)$ scans with heterogenous image contrast due to different stroke phases.

\paragraph{Multi-center data.} 
We test the ensemble algorithm performance over scans acquired in an external imaging center unseen during the development (train phase) of the models. With this aim, 50 test-phase brain MRI scans from center $\#$3 (University Medical Center Hamburg-Eppendorf) are evaluated and compared against 100 scans available during the train phase from centers $\#$1 (University Hospital of the Technical University Munich) and $\#$2 (University Hospital of Bern). 

\paragraph{MRI acquisition time.} 
The ensemble algorithm is evaluated over two sub-groups of the test set data clustered based on the stroke phase. The first group considers scans (n=100) acquired during the late acute or early sub-acute course of the disease. In these cases, MRI is acquired after treating the patient with mechanical thrombectomy. The second group considers patients (n=50) imaged during the early acute phase of the disease and, therefore, MRI is acquired before treating the patient with mechanical thrombectomy.

\paragraph{Lesion size.}
Ischemic stroke spans from minor brain lesions of a few milliliters to large-vessel occlusions involving over a hundred milliliters of brain tissue. Therefore, to understand the algorithm performance when dealing with different ischemic lesion sizes, the test-phase data is split into three stroke sub-groups: lesions smaller than 5 ml, lesions bigger or equal to 5 ml but smaller than 20 ml, and lesions greater or equal than 20 ml.     
\paragraph{Vascular brain territory.}
In this experiment, we evaluate if the deep learning models can identify the affected brain vascular territory in the MRI scans. For doing so, the test-phase scans are linearly registered to a FLAIR MNI template with vascular territory annotations \cite{schirmer2019spatial} using NiftyReg \cite{ourselin2002robust}. Later, the lesion load over each vascular territory is quantified using the ground truth annotations and each scan receives a label of the vascular territory that yields the largest lesion load. The considered vascular territories are the ones covered by the middle cerebral artery, the anterior cerebral artery, the posterior cerebral artery, the arteries perfusing the cerebellum, and the ones perfusing the pons and medulla. The deep learning predictions of the vascular territories are generated by finding the vascular territory with the largest (predicted) lesion volume. Then, we assess through classification metrics how well the algorithms identify the stroke vascular territory.

\paragraph{Stroke pattern.}
The test-phase scans are assigned to one of the three following clinical sub-groups depending on the type of lesion and stroke pattern:
\begin{itemize}
    \item \textbf{No ischemia}\\ Scans with no ischemic lesions (lesion volume of 0 ml, n=2). 
    \item \textbf{Single vessel infarct} \\
    Scans with the largest lesion accounting for >95\% of the total lesion volume (n=62).
    \item \textbf{Scattered infarcts based on micro-occlusions} \\ 
    Scans with $\ge$ three single lesions where either the largest lesion represents < 60$\%$ of the total lesion volume or the total lesion volume is < 5 ml (n=48).
    \item \textbf{Single vessel infarct with accompanying scattered infarcts} \\ 
     All remaining scans (n=38).
\end{itemize}

To label the scans, we perform an iterative computer-human approach as follows. First, using prior knowledge from experienced neuroradiologists (BW and JSK) we define heuristic classification rules that assign each scan to one of the subgroups. Later, the same neuroradiologists evaluated the labels assigned to the scans and updated the heuristics rule, improving its labeling performance. After some iterations, the heuristics rules that suffix the stroke pattern grouping are the ones mentioned above. In order to evaluate if the algorithms can predict the stroke lesion subgroup, these heuristic rules are applied to each (predicted) stroke mask. Then, the stroke subgroup predictions are compared against the `real' labels obtained through the ground truth stroke masks. Conventional classification metrics are used to evaluate the algorithm's performance.

\subsubsection*{Deep learning versus experts in a Turing-like test}
Nine radiologists from three healthcare centers (University Hospital of the Technical University Munich, University Medical Center Hamburg-Eppendorf, and University Hospital of Bern) blindly rated the quality of the ischemic stroke masks generated either by experts or by the ensemble algorithm. Forty or forty-one scans with three annotated slices each (two axial, one sagittal) were provided to each radiologist. All images were randomized, and no information about the annotator (human or algorithm) was provided. Radiologists were asked to rate the \emph{completeness} of the segmented lesion and the \emph{correctness} of their contours on a 1-6 scale as similarly done in Kofler et al.\cite{kofler2021we} (see Supplementary material S6 for the criteria used). 

\subsubsection*{Validating the ensemble algorithm over external data}
The ensemble algorithm is tested over a public, external, ischemic stroke cohort (n = 1686) \cite{liu2023large} including raw MRI (such as DWI, ADC, FLAIR, etc.), patient (e.g., age, sex, race) and clinical data (e.g., treatment, NIHSS and mRS scores, etc.). Table \ref{table: dataset-summary} summarises the dataset characteristics. Images were acquired over ten years using eleven 1.5T or 3T scanners from the four major machine vendors (Siemens, GE, Toshiba, and Philips). NIHSS and mRS scores were respectively performed at patient admission and at 90-day follow-up. Moreover, the time from symptom onset to MRI acquisition was recorded when the patient or the caregiver was confident about symptom onset. MRI was mostly performed six or more hours from symptom onset, before or after administration of intravenous tissue plasminogen activator. To predict ischemic lesions with the ensemble algorithm, all scans were priorly skull-stripped using HD-BET \cite{isensee2019automated}. 

\subsubsection*{Statistical analysis}
\label{section:stats}
Data are compared using the non-parametric, Wilcoxon unpaired rank-sum, or paired signed-rank tests after observing that data is heteroscedastic and does not follow a Gaussian distribution. The significance level is set at $\alpha=0.05$. p-values are adjusted by the Benjamini-Hochberg false discovery rate for multiple comparisons \cite{benjamini1995controlling}. Bland-Altman \cite{bland1986statistical} analysis is used to evaluate the volumetric bias between the manually-traced and the algorithm-predicted lesion volumes. Classification metrics used to evaluate the algorithms are per-class F1 scores ($\mathrm{\text{F1 score}_\text{c} = \frac{2*TP_c}{2*TP_c + FP_c + FN_c}}$) and the balanced accuracy computed as the macro-average of the recall scores per class ($\mathrm{\text{Balanced Accuracy} = \frac{1}{C} \: \sum_{c=1}^{C} \text{Recall}_\text{c}}$, with $\mathrm{\text{Recall}_c= \frac{TP_c}{TP_c + FN_c}}$, TP are true positives, FP the false positives, FN the false negatives and C the number of classes). The scikit-learn Python library \cite{scikit-learn} is used to compute the classification metrics. 

\subsection*{Data availability}
\subsubsection*{Images} The ISLES'22 dataset used for the challenge is open and freely available under the Creative Commons CC BY 4.0 license. The train dataset is available in \url{www.zenodo.org/}\cite{islesdata}. The external stroke dataset used for validating the ensemble algorithm is available through ICPSR \cite{icpsr, liu2023large}.

\subsubsection*{Challenge results} Performance metrics are available in Table \ref{table:ranking} and also through \url{https://grand-challenge.org/}\cite{GC_isles}. Note that this challenge continues accepting submissions and, therefore, the online leaderboard is constantly getting updated. In this study, only the solutions received for the MICCAI'2022 challenge are evaluated.

\subsection*{Code availability}
\subsubsection*{Challenge scripts}
To help participants get familiar with the data and with the challenge performance metrics a Python notebook \cite{islesrepo} was made available in advance. Moreover, in order to facilitate the teams during the algorithms' submission process, a Docker template and a Docker creation tutorial were shared with the teams \cite{dockerisles}. Figures of this work were generated in {\tt R} \cite{computing2013r} using the {\tt ggplot2}\cite{hadley2016ggplot2} and the {\tt patchwork}\cite{patchwork} software packages. Figure \ref{fig:qualitative} was generated with 3D Slicer (\url{https://www.slicer.org/}) \cite{fedorov20123d}. 

\subsubsection*{Ensemble algorithm}
The ensemble algorithm is fully available at \url{https://github.com/Tabrisrei/ISLES22_Ensemble}.

\bibliography{references}

\section*{Acknowledgements}
 PA and SVH received funding from the Flemish Government (AI Research Program) and are affiliated to Leuven.AI - KU Leuven institute for AI, B-3000, Leuven, Belgium. SLL received funding from the National Institutes of Health, National Institutes of Neurological Disorders and Stroke (NIH NINDS), grant R01NS115845. JK receives funding from the Swiss National Science Foundation and the Swiss Heart Foundation outside the submitted work. HA receives funding from the Swiss Heart Foundation outside the submitted work. CH is supported by the Basic Science Research Program through the National Research Foundation of Korea (NRF), funded by the Ministry of Education(2020R1A6A1A03047902). CK is partly supported by the Institute of Information \& communications Technology Planning \& Evaluation (IITP) grant funded by the Korea government (MSIT) (No.2019-0-01906, Artificial Intelligence Graduate School Program(POSTECH)) and Korea Evaluation Institute of Industrial Technology(KEIT), grant funded by the Korea government (MOTIE).

\section*{Author contributions statement}
EdlR, MR, SL, RW, UH, BM, JSK, and BW designed and organized the challenge. EdlR provided support to the challenge participants, evaluated the algorithms, conducted the manuscript experiments, and performed the statistical analysis. EdlR, MR, JSK, BW, MRHP, and SL organized the post-challenge MICCAI workshop.
EdlR, JSK, and BW coordinated the entire work and wrote the document. EdlR, VA, AB, AH, FK, IE, and SS provided technical support to the challenge platform, challenge websites, and/or challenge algorithm testing. AM and JAM supported the algorithmic submissions through \url{https://grand-challenge.org/}. DR, DMS, SW and JK contributed to the conceptual design of the challenge and/or to the post-challenge analysis. SG, KL, ZZ, MMRS, AM, PA, SVH, HJ, CY, CK, JH, SO, RS, AC, AO, XL, LC, IP, JB, EH, JM, NH, CF, AQ, MM, DP, SL, CJ, TH, MG, LB and YL participated in the challenge.
EdlR, RW, UH, WV, MRHP, JSK, and BW contributed to the challenge data selection, organization, and delineation. RW, UH, AH, RZ, GB, CH, MRHP, JSK, and BW performed qualitative ratings for the Turing-like test. 

\section*{Competing interests}
EdlR (DR, VA, DMS, AB) was (are) employed by ico\textbf{metrix}. HA received compensation as a speaker from Bayer AG. CK has financial interests in OPTICHO, which, however, did not support this work.

\end{document}